\pdfoutput=1
\documentclass[aps,prl,reprint,showpacs,superscriptaddress,floatfix]{revtex4-1}

\usepackage{amsmath}
\usepackage{amsfonts}
\usepackage{amssymb}
\usepackage{graphicx}
\usepackage{bm}
\usepackage{braket}
\usepackage{color}
\usepackage{ulem}
\newcommand{\iu}{\mathrm{i}}	%Symbol for the imaginary unit i
\newcommand{\de}{\mathrm{d}}	%Symbol for the differential
\newcommand{\ee}{\mathrm{e}}	%Symbol for the Euler number

\begin{document}

%\preprint{}

\title{The Effect of Multiple Conduction Bands on High Harmonic Emission from Dielectrics}


\author{Peter G. Hawkins}
\email[]{peter.hawkins08@imperial.ac.uk}
\affiliation{Department of Physics, Imperial College London, South Kensington Campus, SW7 2AZ London, United Kingdom}
\author{Misha Yu. Ivanov}
\affiliation{Department of Physics, Imperial College London, South Kensington Campus, SW7 2AZ London, United Kingdom}
\affiliation{Max-Born-Institute, Max-Born Strasse 2A, D-12489 Berlin, Germany}
\author{Vladislav S. Yakovlev}
\affiliation{Max-Planck-Institut f\"{u}r Quantenoptik, Hans-Kopfermann-Stra\ss e 1, D-85748 Garching, Germany}
\affiliation{Department f\"{u}r Physik der Ludwig-Maximilians-Universit\"{a}t M\"{u}nchen, Am Coulombwall 1, D-58748 Garching, Germany}
\affiliation{Center for Nano-Optics, Georgia State University, Atlanta, GA 30303, USA}


\date{\today}

\begin{abstract}
  We find that, for sufficiently strong mid-IR fields, transitions
    between different conduction bands play an important role in the generation of
    high-order harmonics in a dielectric. The transitions make a significant contribution
    to the harmonic signal, and they can create a single effective band for the motion of an
    electron wave packet. We show how high harmonic
    spectra produced during the interaction of ultrashort laser pulses with periodic
    solids provide a spectroscopic tool for understanding the effective band structure
    that controls electron dynamics in these media.
\end{abstract}

%\keywords{}

\maketitle

High-order harmonic generation (HHG) from gas targets is now used as a spectroscopic 
tool for  imaging nuclear (see e.g. \cite{Baker_2006_proton,Lein_2007_molecularimaging,Worner_chemicalreaction_10}) 
and electronic (see e.g.  \cite{smirnova2009_nature,Haessler_Attowavepacket__10,Smirnova_strongfield_09,Mairesse_2010_multichannel})
dynamics on the atomic time- and length scales. 
It is sensitive to various aspects of electronic dynamics, from
attosecond processes in neutral systems \cite{Averbukh_2004,Niikura_2005} to  hole dynamics in ions
\cite{smirnova2009_nature,Haessler_Attowavepacket__10,Smirnova_strongfield_09,Mairesse_2010_multichannel}, correlation-driven channel interaction \cite{Sukiasyan_2010,Morishita_2008,Lin_2010}, and time- and space-resolved information on electronic transitions from different molecular orbitals \cite{Serbinenko_2013,Shafir_2012_tunnellingtime,Shafir_2010_probing}.

We show that HHG spectra from periodic solids give insight into
the effective band structure established by a strong driving  mid-infrared laser field.
Pioneering experiments on high harmonic generation from dielectrics \cite{Ghimire_Ex_12,Schubert_14} stimulated a simple model 
offering semi-classical insight into the underlying physics. In this model (\cite{Ghimire_Ex_12,Ghimire_Th_12}; see also \cite{Mucke_11}) electrons first tunnel from a valence band (VB) to a conduction band (CB) at the maxima of the 
electric field. There, they are driven along the single conduction band by the field. 
The harmonic intensity at frequency $\omega$ is then given by $|\omega J(\omega)|^{2}$, where $J(\omega)$
is the Fourier transform of the current, $j(t)$, in the conduction band, $\varepsilon(k)$. 
Since in this model $j(t) \propto v(t) \propto \de\varepsilon/\de k$, where $v(t)$ is the electron group velocity, 
analysis of the harmonic spectrum can yield information about the band structure ($\de\varepsilon/\de k$). This picture predicts that, when the driving mid-IR laser is sufficiently strong to rapidly accelerate electrons to the edge of the Brillouin zone (BZ), Bragg reflections (Bloch oscillations) within the single band would generate most of the high harmonics.

However, if electrons quickly move past the gap between adjacent CBs, they may undergo an 
interband transition.  In this case, the harmonic signal also comes from coherences 
between all participating bands, including the VB \cite{Golde_nanostructures_08,Schubert_14}. Additionally it is also important to account for the temporal structure of all interband transitions, including the VB to CB transition, see e.g. \cite{subcycletransitions_13}. 
Recent theoretical analysis of HHG in bulk solids by Vampa \textit{et al.}
  \cite{Vampa_14_Th} and Higuchi \textit{et al.} \cite{Stockman_14_SF_bulk} accounted for
  the temporal structure of interband excitations, but as two-band models were used in
  both cases, transitions between conduction bands were not considered.

We show that the inclusion of multiple conduction bands leads to additional
contributions to the high-harmonic signal and that, in spite of the increasing complexity,
the essential information about the motion of electrons in multiple conduction bands
is contained in harmonic spectra. In particular, it reflects the formation of a single, effective CB due to the efficient inter-CB transitions for sufficiently strong driving fields. 

We solve the time-dependent Schr\"odinger equation 
(TDSE) for an electron in a periodic potential:
\begin{equation}\label{fullTDSE}
\hat{H}\Psi(\bm{r},t) = \left[\dfrac{\left[\bm{p} + \bm{A}(t)\right]^{2}}{2} + U(\bm{r}) \right]\Psi = \iu\dfrac{\partial \Psi}{\partial t}  ,
\end{equation}
where $\bm{A}(t)$ is the vector potential of the electric field $\bm{F}(t)=-\bm{A}^{\prime}(t)$, and $U(\bm{r})$ is the periodic potential of the crystal. In \eqref{fullTDSE} and below atomic units are used.
We write  $\Psi(\bm{r},t)$ for an initial crystal momentum $\bm{k}_{0}$ as 
\begin{equation}\label{eq:ansatz}
 \ket{\Psi_{\bm{k}_0}(t)} = \sum_n \alpha_{\bm{k}_0}^n(t)
  \ee^{-\iu \int^t \epsilon^{n}_{\bm{k}(t')}\,\de t^{\prime}} \ee^{-\iu \bm{A}(t) \bm{r}} \ket{\phi_{\bm{k}(t)}^n}  ,
\end{equation}
where $\ket{\phi_{\bm{k}(t)}^n}$ and $\epsilon_{\bm{k}(t)}^n$ are the Bloch states and associated energies of the field-free system, with the time dependence of the crystal momentum being $\bm{k}(t) = \bm{k}_0 + \bm{A}(t)$.
The index $n$ labels the band of the state and $\bm{k}_0$ parametrises the drift momentum. Note that $\ee^{-\iu \bm{A}(t) \bm{r}}\ket{\phi_{\bm{k}(t)}^n}$ are  known as Houston states, see e.g. \cite{Krieger_Iafrate_86}. 
We use the so-called periodic gauge~\cite{Resta_2000_JPCM_12_R107}: $\ket{\phi_{\bm{k}+\bm{G}}^n} = \ket{\phi_{\bm{k}}^n}$,
where $\bm{G}$ is a vector of the reciprocal lattice. That is, whenever $\bm{k}(t)$ lies outside of the first BZ, the periodicity of wave functions with respect to the crystal momentum is assumed.

Our main focus is the modification of the band structure in strong fields, thus we study the single particle response. Substituting the ansatz \eqref{eq:ansatz} into Eq.~\eqref{fullTDSE} yields the set of coupled differential equations for $\alpha_{\bm{k}_0}^n(t)$:
\begin{equation}
\dot{\alpha}_{\bm{k}_{0}}^{n}(t) = -\iu\bm{F}(t)\sum_{n^{\prime}}\bm{\xi}^{n,n^{\prime}}_{\bm{k}(t)}\alpha_{\bm{k}_{0}}^{n^{\prime}}(t)\ee^{\iu\int^{t}\Delta\varepsilon_{\bm{k}(t')}^{n,n^{\prime}} \de t^{\prime}}
\end{equation}
Here, $\Delta\varepsilon_{\bm{k}(t)}^{n,n^{\prime}} = \varepsilon_{\bm{k}(t)}^{n} - \varepsilon_{\bm{k}(t)}^{n^{\prime}}$,
and $\bm{\xi}^{n,n^{\prime}}_{\bm{k}(t)}$ is given by % the Blount matrix of the interband transitions:
\begin{equation}
\bm{\xi}^{n,n^{\prime}}_{\bm{k}(t)} = \iu\braket{\nu^{n}_{\bm{k}(t)}|\nabla_{\bm{k}}|\nu^{n^{\prime}}_{\bm{k}(t)} }  ,
\end{equation}
where $\nu$ is the lattice-periodic part of the Bloch state: 
 $\braket{\bm{r}\vert\phi^{n}_{\bm{k}}}=\nu^{n}_{\bm{k}}(\bm{r})\text{exp}(\iu\bm{k}\bm{r})$. 

After finding $\Psi_{\bm{k}_0}(t)$, we obtain the contributions to the current 
at a particular $\bm{k}_{0}$: $\bm{j}_{\bm{k}_{0}}=\braket{\Psi_{\bm{k}_0}|\hat{\bm{p}}+\bm{A}(t)|\Psi_{\bm{k}_0}}$,
which is then integrated over the BZ to obtain the full current averaged over the unit cell:
\begin{align}
\label{eq:j}
\bm{j}(t) &= \int_{\text{BZ}}\bm{j}_{\bm{k}_{0}}(t) \de^{3}\bm{k}_{0},\\
\label{eq:j_sum}
\bm{j}_{\bm{k}_{0}}(t) &= \sum_{n,n^{\prime}} \bm{a}_{\bm{k}_{0}}^{n,n^{\prime}}(t)
\exp\left(\iu\int^{t}\Delta\varepsilon_{\bm{k}(t')}^{n,n^{\prime}} \de t^{\prime}\right),\\
\bm{a}_{\bm{k}_{0}}^{n,n^{\prime}}(t) &= 
  \left(\alpha_{\bm{k}_{0}}^{n}(t)\right)^{*}\alpha_{\bm{k}_{0}}^{n^{\prime}}(t)\bm{p}^{n,n^{\prime}}_{\bm{k}(t)}.
\end{align}
Here $\bm{p}^{n,n^{\prime}}_{\bm{k}(t)}$ are the momentum matrix elements between Bloch states: $\braket{\phi^{n}_{\bm{k}(t)}\vert\hat{\bm{p}}\vert\phi^{n^{\prime}}_{\bm{k}(t)}}$.

Three distinct physical effects contributing to the generation of high-frequency
components can be identified in the electric current. First, the group velocity, which is
equal to the mean momentum $\bm{p}_{\bm{k}(t)}^{n,n}$, changes its sign as an electron crosses a boundary of
the BZ remaining in the same band, in which case it experiences a Bragg reflection.
This causes a rapid change of the intraband current, which is the part or Eq.~\eqref{eq:j_sum}
with $n=n^{\prime}$:
\begin{equation}
  \bm{j}_{\bm{k}_{0}}^{\mathrm{(IB)}}(t) =
  \sum_{n}\left|\alpha_{\bm{k}_{0}}^{n}(t)\right|^2 \bm{p}^{n,n}_{\bm{k}(t)}.
\end{equation}
Such Bragg reflections are believed to be the main mechanism responsible for the observed
HHG~\cite{Ghimire_Th_12,Schubert_14,Stockman_14_SF_bulk}. Second, the coherent
superposition of any two states with an allowed dipole transition results in quantum
beats. This contribution was analysed in \cite{Vampa_14_Th}, where it was pointed out that
dephasing strongly suppresses the quantum-beat signal, which is dominant
otherwise. Finally, transitions between conduction bands that occur in the regions where
gaps are small can also lead to a very rapid change in the terms associated with interband
coherences ($n \ne n^{\prime}$), provided that $p_{\tilde{\bm{k}}}^{n,n_0} \ne
p_{\tilde{\bm{k}}}^{n^{\prime},n_0}$, where $n_0$ is the index of the electron's initial
(valence) band, and $\tilde{\bm{k}}$ is a crystal momentum where the gap between bands $n$
and $n^{\prime}$ is minimal.  This last contributions has
not yet been studied.

\begin{figure}[b]
 \includegraphics{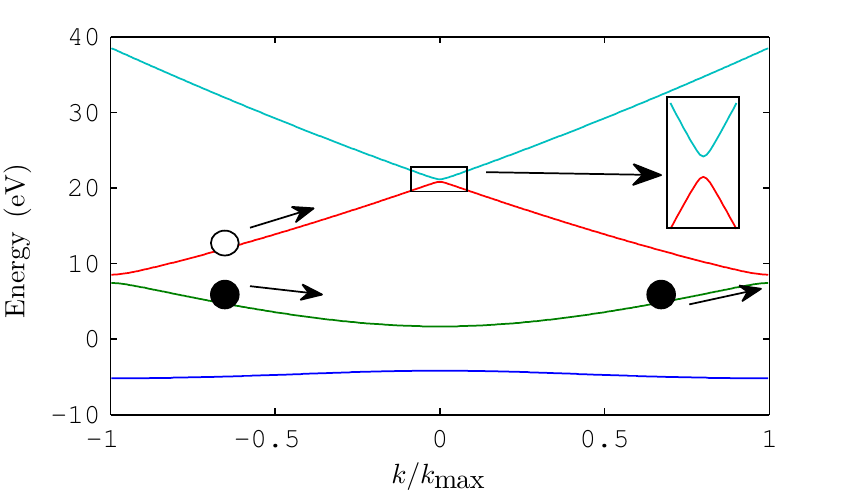}
 \caption{
 The upper valence band and first three conduction bands used in the simulations. Electrons 
 reaching the Bragg plane can stay in the same band, reflecting, shown here as the circle 
 remaining black. Alternatively they can undergo a transition to the next CB, white circle.  
 \label{fig:bands}}
\end{figure}

To obtain an explicit expression for the part of $\bm{j}_{\bm{k}_{0}}(t)$ that
arises as an immediate effect of the external field, we differentiate Eq.~\eqref{eq:j_sum} with
time:
\begin{align}
  \label{eq:dj_dt}
  \frac{\de}{\de t}\bm{j}_{\bm{k}_{0}}(t) &= \frac{\de}{\de t}\bm{j}_{\bm{k}_{0}}^{\mathrm{(tr)}}(t) +
  \frac{\de}{\de t}\bm{j}_{\bm{k}_{0}}^{\mathrm{(QB)}}(t),\\
  \label{eq:djTR_dt}
  \frac{\de}{\de t}\bm{j}_{\bm{k}_{0}}^{\mathrm{(tr)}}(t) &=
  \sum_{n,n^{\prime}} \ee^{\iu\int^{t}\Delta\varepsilon_{\bm{k}(t')}^{n,n^{\prime}} \de t^{\prime}} \frac{\de}{\de t} \bm{a}_{\bm{k}_{0}}^{n,n^{\prime}}(t),\\
  \label{eq:djQB_dt}
  \frac{\de}{\de t}\bm{j}_{\bm{k}_{0}}^{\mathrm{(QB)}}(t) &= \iu \sum_{n,n^{\prime}} \Delta\varepsilon_{\bm{k}(t)}^{n,n^{\prime}}
  \ee^{\iu\int^{t}\Delta\varepsilon_{\bm{k}(t')}^{n,n^{\prime}} \de t^{\prime}} \bm{a}_{\bm{k}_{0}}^{n,n^{\prime}}(t).
\end{align}
In our model, which does not explicitly account for dephasing, the
  quantum-beat current $\bm{j}_{\bm{k}_{0}}^{\mathrm{(QB)}}(t)$ gradually grows as the
  concentration of charge carriers increases, and it persists after the laser pulse. In
  contrast, the derivative of the \textit{transient current} $\bm{j}_{\bm{k}_{0}}^{\mathrm{(tr)}}(t)$
  becomes zero as soon as the external field disappears, and it is affected by any rapid
  change of matrix elements or probability amplitudes that may occur at an avoided crossing between bands. In the following, we
  will focus on $\bm{j}_{\bm{k}_{0}}^{\mathrm{(tr)}}(t)$, assuming that the
  contribution from $\bm{j}_{\bm{k}_{0}}^{\mathrm{(QB)}}(t)$ to sufficiently high
  frequencies is suppressed by dephasing phenomena. This division of the current density
  into two different parts is different from the division in the
  interband and intraband currents in~\cite{Schubert_14,Vampa_14_Th}, while the intraband
  current is fully included in $\bm{j}_{\bm{k}_{0}}^{\mathrm{(tr)}}(t)$. The proposed
  separation of the currents has the drawback that neither
  $\bm{j}_{\bm{k}_{0}}^{\mathrm{(tr)}}(t)$ nor $\bm{j}_{\bm{k}_{0}}^{\mathrm{(QB)}}(t)$
  alone account for the linear polarisation response, but, as we show below, it is very
  useful to analyse and visualise the high-frequency response.

\begin{figure}[t]
\includegraphics{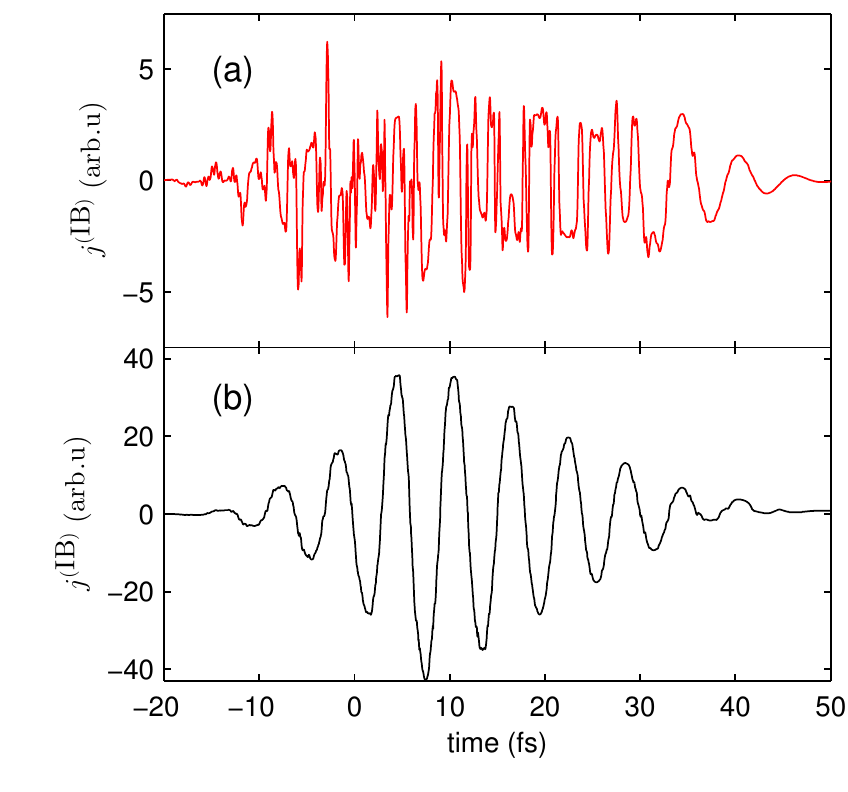}
\caption{Plot (a) shows the intraband current with only 1 CB, in this case interband transition to 
higher CBs cannot happen and so Bragg reflections are forced to happen, explaining the appearance of 
higher harmonic content. Plot (b) shows the intraband current for 8 CBs. \label{fig:j_diag_compare}}
\end{figure}

To simplify our simulations and the subsequent analysis, we solve the
TDSE problem in one spatial dimension. We obtain the energies and matrix
  elements by solving the stationary Schr\"odinger equation for a periodic lattice
  potential $U(x)$ that, within the central unit cell, has the form: $U(x) = -U_0(1+\tanh(x+x_0))(1+\tanh(-x+x_0))$.
This potential allows us to reproduce the key parameter
  of a real solid: the band gap. We chose
  our parameters to model aluminium nitride by assuming a lattice spacing
  of $a=8.15\,\mbox{au}$ and setting $U_0=0.78$, $x_0=0.565$; this yields
  a VB-CB band gap of 5.85\,eV. The gaps 
between the conduction bands are smaller, with the first CB gap being 1.09\,eV. The highest 
VB and first three CBs are shown in Fig.~\ref{fig:bands}, plotted as a function of $k/k_{\text{max}}$ 
with $k_{\text{max}}= \pi/a$.

We consider laser pulses at $\lambda_\mathrm{L}=1800\,\mbox{nm}$
($\omega_\mathrm{L}=2 \pi c / \lambda_\mathrm{L}=1.05\,\mbox{fs}^{-1}$), with a field strength of 0.75V\AA$^{-1}$, and a full width half maximum (FWHM) of 30fs. The envelope used for the vector potential is of the form: $\cos^{4}(\pi t/2\tau)$, where $\tau=(\pi/4)(\tau_{\text{FWHM}}/\cos^{-1}(2^{-0.125}))$.

\begin{figure}[t]
\includegraphics{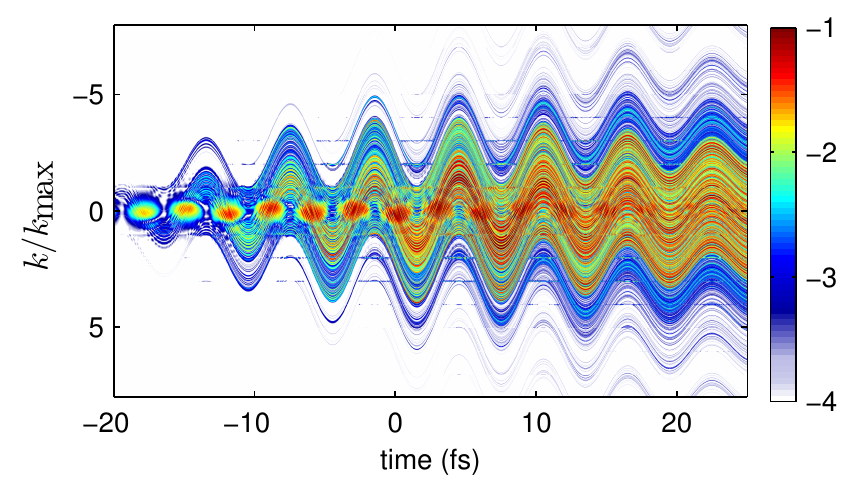}[b]
\caption{
The time dependent conduction band population is plotted on a logarithmic scale in the extended zone 
scheme, so that the n\textsuperscript{th} CB occupies crystal momenta $n-1<|k|/k_\mathrm{max}<n$. The 
high transition probability between CBs is easily seen.
\label{fig:tk_plot}}
\end{figure} 

To highlight the importance of multiple conduction bands, as well as the interplay of Bragg reflections 
and transitions between conduction bands, in Fig.~\ref{fig:j_diag_compare} we show the intraband 
current. For a single CB, the intraband current shows strong 
Bragg reflections and Bloch oscillations, Fig.~\ref{fig:j_diag_compare}(a). 
However, as soon as multiple bands are included, these effects disappear and 
the intraband current is dominated by the fundamental frequency,  Fig.~\ref{fig:j_diag_compare}(b). 
Clearly, for such fields a simulation  with a single CB is inadequate, we used 8 CBs as this is the requirement for convergence. The dominance of the 
fundamental implies that electrons are moving on 
a single effective parabolic potential, reflecting dramatic modification 
of the band structure due to the dominance of the interband transitions over Bragg reflections. 

This effect is easily visualised by plotting $|\alpha^{n}_{\bm{k}_{0}}(t)|^{2}$ for conduction bands
in the extended zone scheme, as can be seen in Fig.~\ref{fig:tk_plot}. As electrons pass by BZ edges 
the most probable path changes from Bragg reflection early in the pulse into transition into the higher CBs as the IR intensity grows.

\begin{figure}
\includegraphics{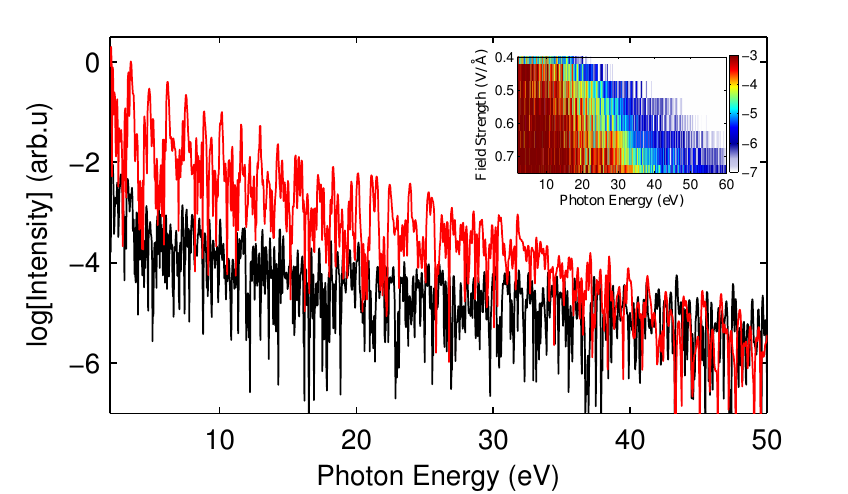}
\caption{
The harmonic spectrum generated by the transient current is plotted for the single and many 
CB cases. The variation of the spectra with field strength is shown in the inset plot for the single CB case.
 \label{fig:transient_spectra}}
\end{figure}

\begin{figure}[b]
\includegraphics{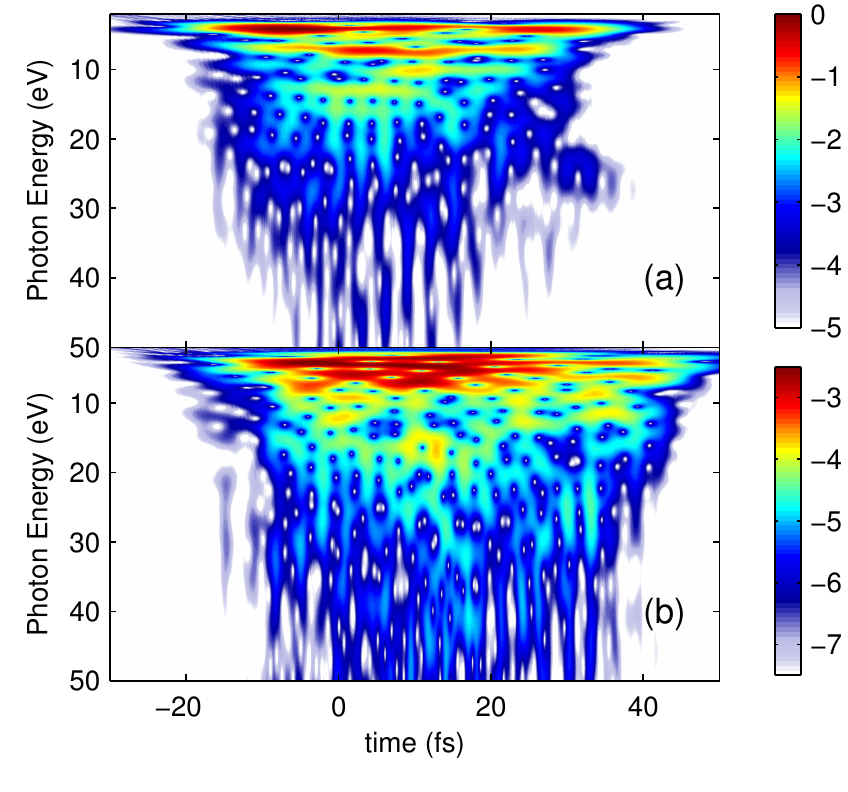}
\caption{Time-frequency analysis, as described in the text, of the transient current for the single 
(a) and many(b) CB cases is plotted. One can immediately see that the emission for the single CB case 
is occurring in bursts that are more defined than in the many band CB case.
 \label{fig:hhg_wavelet}}
\end{figure} 

\begin{figure}[b]
\includegraphics{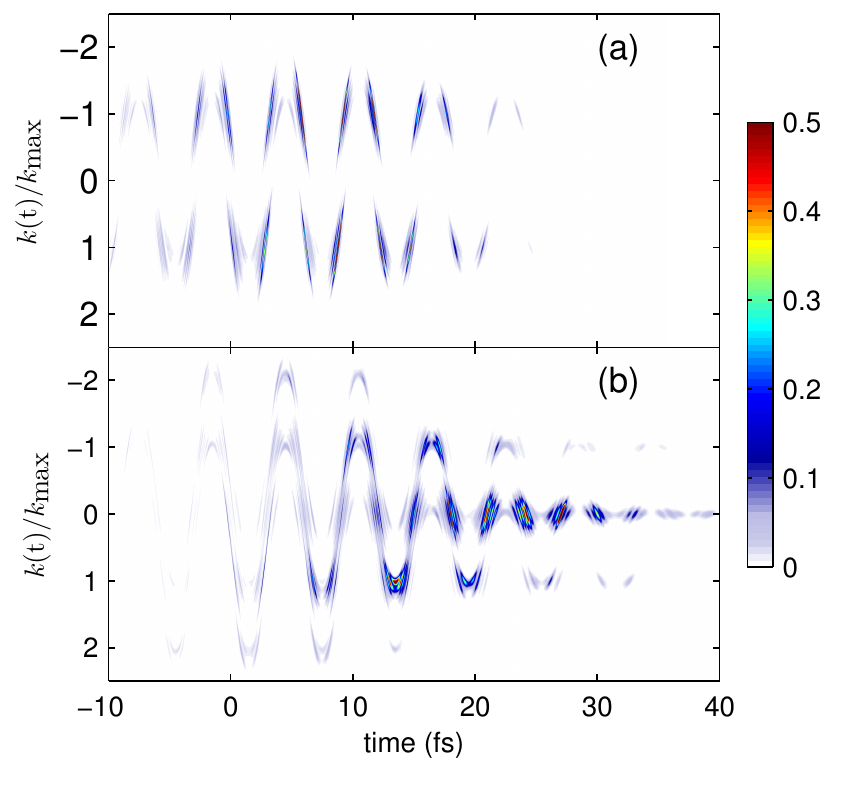}
\caption{The strength of the harmonic emission given by the transient current, when multiplied with 
a Gaussian window centred around 30eV and FWHM of 5eV, is plotted as a function of the time dependent 
crystal momentum and time for the single (a) and many (b) CB cases. For the single band case we see the 
emission is occuring at the Bragg planes: $|k|=k_\mathrm{max}$. For the many-band case a lot of the emission is at $k=0$, 
particularly at larger times.  \label{fig:spectral_tk}}
\end{figure}

We now focus on the transient current \eqref{eq:djTR_dt}. In Fig.~\ref{fig:transient_spectra} 
the spectrum of the transient current is shown for the many and single CB cases. The harmonic content is 
much more defined for the single CB case. Inset in Fig.~\ref{fig:transient_spectra} is the field dependence 
of the spectra for the single CB case, in a similar style of plotting to that used in \cite{Stockman_14_SF_bulk}. 
The cutoff scaling is seen to be linear with field strength as in experiments \cite{Ghimire_Ex_12,Schubert_14}. 
Note that for the multiple CB case with the band structure used here, transitions saturate quickly leading to a breakdown 
of the cutoff scaling.  

To understand the time dependence of the harmonic emission we employ time-frequency analysis of the 
transient current obtained via the Morlet wavelet transform:
\begin{equation}
W(\Omega,\tau) =  \int\,\de t\,j^{\text{(tr)}}(t)\ee^{\iu\Omega t}\ee^{-\left(\dfrac{\Omega(t-\tau)}{\sqrt{2}\sigma\Omega_c}\right)^2},
\end{equation}
where $\sigma$ is selected to yield 14\% of a cycle width at $\Omega_c=15 \omega_\mathrm{L}$. 
The width of the time--domain window then decreases with increasing $\Omega$, improving the resolution.  

The result is plotted in Fig.~\ref{fig:hhg_wavelet}.  For the single CB case, emission is 
half-cycle periodic, with bursts around the peaks of the electric field. This is particularly true for photon 
energies above the maximal bandgap between the upper VB and lower CB. Indeed, this is when the electrons experience highest acceleration
past the BZ edge, thus generating the highest harmonic content when Bragg reflected in the single-CB model. 
However when multiple CBs are included the bursts of emission is not so well defined temporally. We also
see that the most intense harmonic emission occurs after the centre of the pulse in both cases. This is because the 
concentration of charge carriers continues to grow after the peak of the pulse, which compensates for the decrease of the field.

To see where in the band structure and at what times in the field harmonic emission occurs, we develop another technique. The 
harmonic spectrum generated by electrons with initial crystal momentum $\bm{k}_0$ is given by the Fourier transform of $\bm{j}_{\bm{k}_{0}}^{\mathrm{(tr)}}(t)$.
We take the product of a Gaussian window with 
the harmonic spectrum for a given $\bm{k}_0$ to select a spectral region of interest: 
\begin{equation}
J^{(\text{tr})}_{\bm{k}_{0}}(\omega ;\omega_0 , \sigma) =  \mathcal{F} \left[ \bm{j}_{\bm{k}_{0}}^{\mathrm{(tr)}}(t)\right]
\exp\left(-\dfrac{(\omega-\omega_0)^2}{2 \sigma^2}\right).
\end{equation}
This allows us to investigate the temporal profile of emission in this spectral region: for a given $\bm{k}_0$, the envelope of harmonic bursts 
is thus given by $E(t,\bm{k}_0 ;\omega_0 , \sigma) = \mathcal{F}^{-1} \left[J^{\text{tr}}_{\bm{k}_{0}} \right]$. Since 
every $\bm{k}_0$ is related to $\bm{k}(t)$ it also allows us to map the harmonic emission in the spectral region to the 
time at which it occurs, and the crystal momentum at that time. 

In Fig.~\ref{fig:spectral_tk} this analysis is applied to compare the nature of the harmonic emission in 
the single and many CB cases. It 
allows us to clearly see that for the single CB case the harmonic emission is dominated by electrons reflected 
at Bragg planes. For the many CB case the process is modified, we still see that there is some emission around 
the Bragg plane from electrons reflecting at and crossing it. At later times there is emission for electrons 
with crystal momentum of $k=0$, this acts to prolong the time over which emission occurs compared to the single CB case. We attribute the prolonged emission at $k=0$ to be due to electrons crossing between the second and third CBs, where the gap is small (see Fig.~\ref{fig:bands}) so that although transition probability even at reduced field is still large. 

It was observed in experiments that HHG in a crystalline solid is particularly efficient for certain orientations of the sample with respect to the polarisation of the fundamental field \citep{Ghimire_Ex_12,Schubert_14}. Our results suggest that the observed angular dependence is mainly due to transitions between CBs, which occur at local minima of inter-CB energy gaps, and the probability of which is very sensitive to the magnitudes of the gaps. 

Our most important finding is that such interband transitions not only reduce the intensity of harmonic emission, but they can also have a strong impact on its spectral and temporal properties: individual harmonics become less distinct, and the rapid change of the quantum-beat signal associated with interband transitions plays a particularly important role in higher CBs. To study these effects, we have identified a useful quantity, the transient current, that allows the nature of the harmonic emission to be disentangled from quantum beats that are expected to be strongly suppressed by dephasing in real solids. The relative importance of Bragg reflections and interband transitions is sensitive to the band structure and field parameters, but the statement that Zener-like transitions between conduction bands result in the emission of high-frequency radiation is general. This effect may be used to experimentally study the motion of electrons driven by a strong mid-IR field in an effective single nearly parabolic band, once temporal characterisation of harmonics emitted from a solid sample becomes feasible.

\begin{acknowledgments}
P.H. and M.I. were supported by EPSRC programme grant EP/I032517/1, and also acknowledge support from Marie Curie ITN CORINF. V.S.Y was supported by the DFG Cluster of Excellence: Munich-Centre for Advanced Photonics (MAP), he is indebted to S. Kruchinin, E. Goulielmakis and M. Stockman for useful discussions. This work was (partially) supported by The United States Air Force Office of Scientific Research under program No. FA9550-12-1-0482.
\end{acknowledgments}

%\clearpage
\bibliography{manybands_refs}

\end{document}